\documentclass[twoside,12pt]{article}
\usepackage{CJK}
\usepackage{indentfirst}
\usepackage{bm}
\usepackage{epsfig}
\usepackage{color}
\usepackage[normalem]{ulem}

\usepackage{graphicx}
\usepackage{physics}
\usepackage{caption}
\usepackage{subcaption}
\usepackage{amssymb}

\begin{document}

\begin{CJK*}{GBK}{song}

%-------------------  First Head  -----------------------------------------
\thispagestyle{empty} \vspace*{0.8cm}\hbox
to\textwidth{\vbox{\hfill\huge\sf Commun. Theor. Phys.\hfill}}
\par\noindent\rule[3mm]{\textwidth}{0.2pt}\hspace*{-\textwidth}\noindent
\rule[2.5mm]{\textwidth}{0.2pt}

%=================== Textbegin here =============================================

\begin{center}
\LARGE\bf {Quantum Monte Carlo study\\ of the superfluid phase of $^4$He}
\end{center}

\footnotetext{\hspace*{-.45cm}\footnotesize $^\dag$Corresponding author, E-mail: m.boninsegni@ualberta.ca }

\begin{center}
\rm M. Boninsegni$^\dagger$
\end{center}

\begin{center}
\begin{footnotesize} \sl
Department of Physics, University of Alberta, Edmonton, Alberta, Canada T6G 2H5\\
Phenikaa Institute for Advanced Study, Phenikaa University Nguyen Trac Street,\\ Duong Noi Ward, Hanoi, Vietnam
\end{footnotesize}
\end{center}

\begin{center}
%\footnotesize (Received 2 March 2024; revised manuscript received 16 May 2024)

\end{center}

\vspace*{2mm}

\begin{center}
\begin{minipage}{15.5cm}
\parindent 20pt\footnotesize
Detailed numerical results obtained with state-of-the-art Quantum Monte Carlo (QMC) simulations are presented for the superfluid phase of $^4$He at saturated vapor pressure. The aim of this contribution is that of providing reliable, up-to-date estimates for this archetypal superfluid, reflecting the methodological progress that has taken place over the past two decades. We simulate a system comprising 2,048 helium atoms, i.e., an order of magnitude greater in size than those for which results currently regarded as standard references were originally obtained. We offer revised estimates for energetic and structural properties, as well as for the ground state condensate fraction.
\end{minipage}
\end{center}
\bigskip\bigskip
\begin{center}
\begin{minipage}{15.5cm}
\begin{minipage}[t]{2.3cm}{\bf Keywords:}\end{minipage}
\begin{minipage}[t]{13.1cm}
Quantum fluids; Quantum Monte Carlo; Superfluid $^4$He
\end{minipage}\par\vglue8pt

\end{minipage}
\end{center}
\newpage
\section{Introduction}
Forty years after the pioneering work of Ceperley and Pollock \cite{Ceperley1986,Pollock1987}, showing how the superfluid transition of $^4$He could in principle be studied by computer simulation, QMC methods based on Feynman's space-time formulation of quantum statistical mechanics \cite{Feynman1965} are now accepted as a legitimate, powerful and often the best option to compute accurate thermodynamics of strongly interacting Bose systems, essentially with no approximation. In particular, following the introduction two decades ago \cite{Boninsegni2006,Boninsegni2006b} of the continuous-space Worm Algorithm (WA), overcoming the limitations of the original Path Integral Monte Carlo methodology, QMC simulations at finite temperature have been utilized to investigate a variety of bulk superfluid phases of Bose systems in continuous space, featuring, among others, soft-core \cite{Saccani2011,Boninsegni2012c} dipolar \cite{Kora2019,Filinov2010} and Coulomb \cite{Zhang2023,Boninsegni2026} interactions among the elementary constituents. 
\\ \indent 
Surprisingly, current results for the superfluid phase of $^4$He (summarized in Ref. \cite{Ceperley1995}) date back a few decades and have remained largely unrevised, with the exception of the superfluid response \cite{Boninsegni2006b}.  Besides being affected by statistical uncertainties that are objectively large by today's standards, those results were obtained by simulating systems of rather small size (typically $\sim$ 100 atoms); thus, they are generally not reliably representative of the thermodynamic limit. 
They are also rather incomplete; for example, no (reliable) estimates for the energetics and the condensate fraction at finite temperature are available in the limit in which the ground state is approached. One may wonder why no effort has gone into generating updated results, by taking advantage of ever improving computing facilities. After all, this would seem a worthwhile goal, considering that $^4$He is not only  an important physical system in its own right, but also the simplest non-trivial superfluid, for which a wealth of very accurate experimental measurements have been performed \cite{Barenghi1998}, allowing one to gauge the accuracy of the most advanced microscopic models and computational methods. Moreover, because QMC allows one to carry out essentially exact studies of interacting Bose systems, accurate simulation data can serve to benchmark novel theoretical approaches.  \\ \indent 
Part of the answer is undoubtedly that the original methodology  \cite{Ceperley1995} failed to deliver upon its initially stated promise, due to inefficient sampling and ensuing poor scaling {\em vis-\`a-vis} the number of particles in the simulated system. Indeed, to this day no QMC study of bulk superfluid $^4$He based on that simulation scheme and featuring more than $\sim 100$ particles \cite{Boninsegni1997} has been carried out. Only after the introduction of the continuous-space WA in 2006 has it become possible to obtain by simulation reliable numerical estimates of cogent physical quantities such as the superfluid transition temperature, which requires accurate finite-size scaling analysis, i.e., simulations on systems of {\em a}) sufficiently large and {\em b}) significantly different size (e.g., a few thousand particles).
More generally, however, superfluid $^4$He has been considered for the most part a ``solved problem'', no longer at the forefront of theoretical research, and attention has shifted to other systems (e.g., cold atom assemblies).
\\ \indent
The purpose of this contribution is to provide a fairly comprehensive set of up-to-date QMC data for energetic, structural and superfluid properties of the bulk equilibrium phase of condensed $^4$He at low temperature ($0.25$ K $ \le T \le$ 2 K).
This study complements and extends the work of Ref. \cite{Boninsegni2006b}, mainly by  providing estimates for energetics, structural correlations, superfluid and condensate fractions as a function of temperature for bulk three-dimensional $^4$He, also assessing the dependence of the results on the pair potential utilized to model the interaction among two helium atoms. Another, broader aim is to clarify some methodological aspects about which some confusion (and even actual misconceptions) may still exist, especially among non-practitioners of QMC. 
\\ \indent
This paper is organized as follows: the remainder of this section briefly rehashes some basic information about superfluid $^4$He, to clarify the terminology and facilitate the reading to a non-specialist. The microscopic model of the system is described in Sec.\ \ref{model} \ \ , while the computational technique utilized is briefly reviewed in Sec.\ \ref{meth}\ \ \ ; results are presented in  Sec.\ \ref{res}\ \ \ \ and discussed in Sec.\ \ref{concl}\ \ \ 
\subsection*{Background}
The naturally occurring condensed phase of $^4$He is a remarkably close physical realization of a simple model of strongly interacting Bose system. In the pressure and temperature range in which superfluid helium exists,  helium atoms can be regarded to an excellent approximation as ``elementary constituents'', i.e., individual ``point-like'' particles of spin zero (hence obeying Bose statistics). 
Furthermore, the potential energy of interaction is very nearly given by the sum of terms associated to pairs of atoms, each term described by a potential energy function only depending on the relative distance between the two atoms. The main feature of such a potential is the presence of a strong repulsive core at short interparticle separation ($\le$ 2.6 \AA), whose physical origin lies in the Pauli repulsion among electrons of different atomic clouds. Indeed, in his seminal papers on this subject Feynman posited that treating schematically the interaction between two helium atoms as an infinitely repulsive wall at short distance (and zero everywhere else) may be adequate, in order to gain at least qualitative insight into the thermodynamic properties of the system \cite{Feynman1953}.
\\ \indent
Experimentally, superfluid $^4$He is one of the most extensively investigated condensed matter systems. Its behavior at low temperature is unique and peculiar. First, it fails to crystallize at low temperature under the pressure of its own vapor, remaining a liquid all the way to temperature $T=0$.  Second, at temperature $T=2.178$ K it undergoes a transition to a superfluid phase, capable of flowing without dissipation through very narrow capillaries. These two properties can be exploited to purify the system to a very high degree, in turn allowing for highly controlled and precise experiments \cite{Wilks1967}.
\\ \indent
As condensed $^4$He can be regarded as the archetypal quantum (Bose) fluid, and because a reliable comparison of theoretical predictions and experimental observations is possible (perhaps to a greater degree than in any other known condensed matter systems), the ambitious theoretical goal consists of achieving an accurate microscopic description, based on realistic interatomic potentials, capable of delivering reliable predictions for experimentally relevant quantities. Thus, starting from a many-body Hamiltonian in which the only external (independent) input is the potential energy of interaction among atoms, one aims at carrying out calculations based on fundamental quantum mechanical equations without resorting to any uncontrolled approximation.
It seems fair to state that QMC simulations at finite temperature have achieved such a goal to a remarkable degree, and played a major role in shaping our current theoretical understanding of the superfluid phase of $^4$He, and more generally of superfluidity and its connection to Bose condensation. The results offered here are meant to provide quantitative substantiation to such a claim, by extending numerical predictions to lower temperatures and providing a more precise, robust and definitive foundation to the existing quantitative picture.

\section{Model}\label{model}
As mentioned above, the system is described as an ensemble of $N$ point-like, identical particles with spin zero and a mass $m$ equal to that of a $^4$He atom. The system is formally enclosed in a cubic cell of volume $\Omega=L^3$, so that $n=N/\Omega$ is the nominal density. Periodic boundary conditions in all directions are  assumed. 
\\ \indent
Regarding helium atoms as point particles means making the Born-Oppenheimer approximation, which is ubiquitously made in theoretical studies of superfluid helium. It is amply justified at conditions of temperature and pressure where the superfluid phase of $^4$He exists, owing to the large (at least three orders of magnitude) difference between between electronic and ionic energy scales. 
The standard, non-relativistic quantum-mechanical many-body Hamiltonian of the system reads as follows:
\begin{eqnarray}\label{u}
\hat H = - \lambda \sum_{i}\nabla^2_{i}+\sum_{i<j}v(r_{ij})
\end{eqnarray}
where the first (second) sum runs over all particles (pairs of particles), $\lambda\equiv\hbar^2/2m=6.0596415$ K\AA$^{2}$, $r_{ij}\equiv |{\bf r}_i-{\bf r}_j|$ and $v(r)$ is the pair potential which describes the interaction between two helium atoms. 
Several different forms for $v$ have been proposed and used in the literature, often complicating a bit the comparison of different calculations. 
Early QMC studies of the condensed phase of $^4$He were based on the simple Lennard-Jones potential \cite{Kalos1974}, whose quantitative shortcomings (e.g., its unphysical ``stiffness'' at short interparticle separation) were quickly recognized, prompting its replacement with more sophisticated model interactions. In particular, the Aziz pair potential introduced in 1979 (henceforth referred to as Aziz I) \cite{Aziz1979} has been shown to provide a reasonably accurate quantitative description of the superfluid phase of $^4$He \cite{Ceperley1995}.
\\ \indent
The Aziz I potential is mainly phenomenological, i.e., adjusted in part to ensure consistency with known aspects of the interaction and basic experimental observations. It effectively incorporates potential energy contributions beyond the pairwise decomposition (e.g., three-body terms). Subsequent versions of this interaction have aimed at being increasingly grounded in first principle quantum chemistry calculations, while retaining the same basic analytical structure; as a result, the agreement between theory and experiment afforded by these refined potentials is typically slightly worse than that achieved with the Aziz I potential. 
The agreement can be improved by the explicit inclusion of three-body terms \cite{Cencek2007,Sese2020}, whose quantitative effect on the physical quantities of interest here is relatively small. Specifically, one observes a rather tiny shift in the potential energy and in the pressure, bringing them in closer agreement with experiment. However, little or no change in the kinetic energy, or in structural or superfluid properties is brought about by the inclusion of three-body terms \cite{Moroni2000,Chang2001}.
For the sake of consistency with the majority of previous studies, the results provided here are obtained with the Aziz I potential, and its use can be assumed unless otherwise stated. Some of the results are compared with those obtained with a later version thereof, proposed in 1995 \cite{Aziz1995} (Aziz II), as well as with the Lennard-Jones potential with standard values of the parameters for He (i.e., $\epsilon=10.22$ K and $\sigma=2.556$ \AA).
\section{Methodology}\label{meth} 
The results described here were obtained by QMC simulations of  the system described by Eq.  (\ref{u}), based on the canonical version \cite{Mezzacapo2006,Mezzacapo2007} of the continuous-space WA. Our calculations are at fixed density, and we utilize published experimental data for the equilibrium density \cite{Barenghi1998}. Arguably the most crucial advantages of the WA over the previously existing prescription are, as mentioned above, its scalability versus system size and the efficient way in which exchanges of indistinguishable particles are sampled, allowing it to simulate systems of much greater size. It is known that quantum-mechanical exchanges, which in Feynman's path integral
language result in the ``entanglement'' of the many-particle paths in imaginary time, underlie collective phenomena such as superfluidity and Bose Condensation  \cite{Ceperley1995,Feynman1965}; they are also responsible for the stability of the liquid phase of $^4$He at low temperature \cite{Boninsegni2012d}.\\ \indent
For a comprehensive description of the WA the reader is referred to Ref. \cite{Boninsegni2006b}; here, besides providing all the relevant technical details required to reproduce the results, some specific computational issues are addressed for which, in our view, the prevailing beliefs and conventional wisdom may be  inaccurate or misleading. 
%\sout{It should be mentioned that in this work the trick known as Diagrammatic 
%Monte Carlo, utilized in Ref. \cite{Boninsegni2006b} to treat more efficiently 
%the interaction among atoms, was not used. In other words, the potential energy 
%of interaction among all pairs of atoms was computed explicitly. This choice was 
%made to reduce computational complexity, and to facilitate the comparison of our 
%results with existing ones (i.e., in the $T\to 0$ limit); clearly, a cost is paid 
%in terms of efficiency, but it is one easily absorbed by  the computing facilities 
%available for this project.}
\\ \indent
Other details of the simulation are standard. 
Physical quantities of interest calculated in this study include, besides pressure and energetics, the pair correlation function and the related, experimentally accessible static structure factor, as well as the superfluid and condensate fractions. All of the estimators used in this work are the ones commonly utilized \cite{Ceperley1995}. For the energy, the thermodynamic estimator has been 
{\color{black} used.}
%{\sout {consistently found to be the most accurate and reliable, unsubstantiated 
%claims published elsewhere of alleged superiority of the virial estimator 
%notwithstanding}}. 
The virial estimator was instead used to calculate the pressure; the superfluid fraction is computed using the well-known winding number estimator \cite{Pollock1987} while the condensate fraction is inferred from the long-distance behavior of the one-particle density matrix $n({\bf r})$, 
{\color{black}defined as 
\begin{equation}
    n({\bf r})={\Omega}^{-1} \int d^3r^\prime\ \langle \hat\psi^\dagger({\bf r}+{\bf r}^\prime) \ \hat\psi({\bf r}^\prime)\rangle
\end{equation}
where $\langle ...\rangle$ stands for thermal average and $\hat\psi$, $\hat\psi^\dagger$ are the usual Bose field operators. For a homogeneous, isotropic fluid it is $n({\bf r})\equiv n(r)$.
The quantity $n({\bf r})$} is a direct outcome of a QMC simulation based on the WA  \cite{Boninsegni2006b}. This is illustrated in the next sections.
\\ \indent
The data quoted in this work are the results of simulations  of systems  comprising $N=2048$ $^4$He atoms. To what a degree the results of a simulation on a system of this size can be regarded as quantitatively representative of the thermodynamic limit is discussed in detail below; with the exception of the superfluid and condensate fraction, for which finite-size corrections remain significant in the vicinity of the transition \cite{Boninsegni2006b}, in our submission the system size considered here gives numerical values indistinguishable from estimates pertaining to the thermodynamic limit, within their statistical uncertainties. For the calculation of the (potential) energy per particle and the pressure, the contribution of particles outside the main simulation cell was estimated through appropriate volume integrations of the pair potential outside the cell. This is a reliable estimation, as the pair correlation function $g(r)$ converges to its asymptotic value of 1 well within the simulation cell, in the density range of interest here, thanks to the relatively large size of the simulated system.
\subsection*{Time step extrapolation}\label{tse}

\begin{figure}[h]
\centering
     \includegraphics[width=0.7\linewidth]{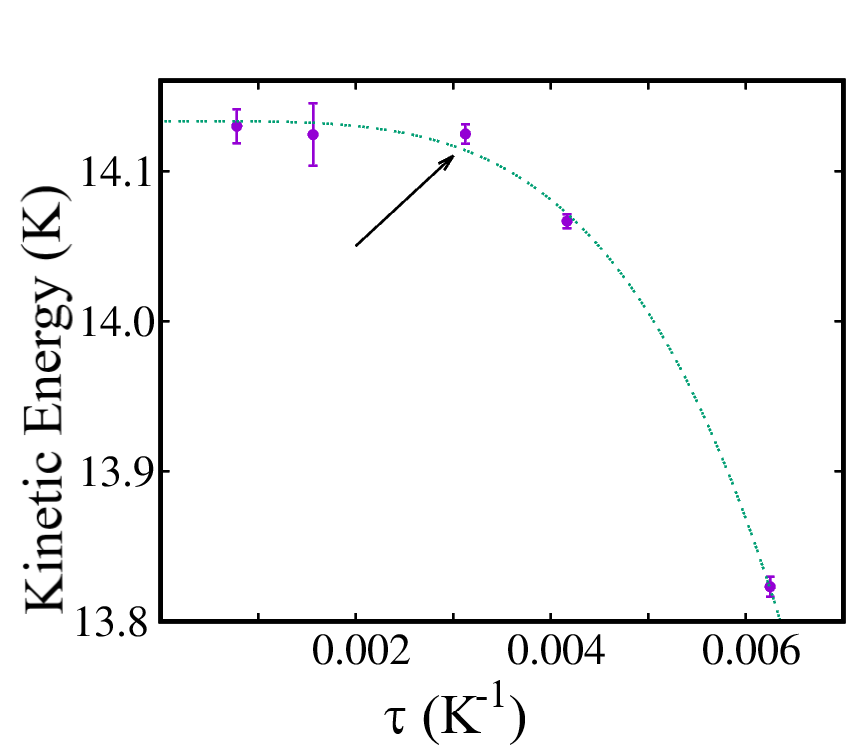}
\caption{Time step extrapolation of the kinetic energy per particle in bulk superfluid $^4$He, at temperature $T=1$ K, for a system comprising $N=2048$ atoms. The density is $0.021834$ \AA$^{-3}$. The results are obtained with the Aziz I pair potential.
The dashed line is  a quartic fit to the data. Arrow points to the value of $\tau$ for which convergence is reached, within statistical uncertainty. The value of the kinetic energy extrapolated to the $\tau=0$ limit is 14.133(5) K. \label{kextrap}}
\end{figure}  

In this subsection we discuss the extrapolation of the physical estimates to the limit of vanishing imaginary time step, an extrapolation that is always implicit in QMC simulations of continuous systems due to the discretization of many-particle path (for details, see Ref. \cite{Ceperley1995}). Specifically, we wish to address a specific aspect not discussed much (if at all) in the literature, but is potentially relevant, as often times the energy is not the main quantity of interest.  
\\ \indent 
For illustrative purposes, we consider the bulk equilibrium phase of $^4$He at temperature $T=1$ K. The density is the experimentally determined equilibrium density $0.021834$ \AA$^{-3}$ \cite{Barenghi1998}. The short-time approximation to the imaginary-time propagator used to perform the calculations whose results are presented here is accurate to fourth order in the time step $\tau$ (see, for instance, Ref. \cite{Boninsegni2005}). This particular form of short-time propagator has been the standard choice for this kind of simulations since the introduction of the WA.\\ \indent
Fig. \ref{kextrap} shows the extrapolation of the estimates of the kinetic energy per $^4$He atom, computed with different time steps. The data, plotted versus $\tau$ (in K$^{-1}$), can be fitted with the functional form $f(\tau)=k_0+k_1\tau^4$, where $k_0,\ k_1$ are fitting parameters; specifically, $k_0=14.133(5)$ K is the value of the kinetic energy per $^4$He atom extrapolated to $\tau=0$. 
\\ \indent
The extrapolation procedure illustrated in Fig. \ref{kextrap} can be computationally expensive, especially if (as it is usually the case) one is interested in carrying out many simulations, e.g., for different thermodynamic conditions. The practical solution consists of identifying a value of time step ($\tau_c$) yielding estimates sufficiently close (i.e., within error bars) to those extrapolated to the $\tau=0$ limit, and carry out a single simulation for every thermodynamic point of interest, using that value of $\tau$. 
For example, looking at Fig. \ref{kextrap} one can see that kinetic energy values obtained with a time step $\tau\le 3.125\times 10^{-3}$ K$^{-1}$, 
all bracket the extrapolated result, within its statistical uncertainty. 
\\ \indent 
What is ubiquitously done, in QMC studies in which a time step extrapolation is required, is make the operative assumption that calculations carried out with the ``optimal'' time step ($\tau_c=3.125$ K$^{-1}$, in the example of Fig. \ref{kextrap}), at which one arrives by ensuring convergence of the (kinetic) energy \cite{comment},   should yield results reliably representative of the $\tau=0$ limit not just for the energy but for {all} physical quantities of interest. This work is no exception, as unless otherwise stated, all the values reported here were obtained using a time step $\tau=\tau_c$.
\\ \indent
There are theoretical reasons to expect the kinetic energy, which is essentially a measurement of the curvature of the paths in imaginary time, should be the most sensitive to time discretization \cite{Ceperley1995}.
It turns out, however, that for the {\em vast} majority of important physical quantities convergence is typically observed with a {\em significantly} greater (as much as a few times) value of $\tau$, which in turn results in a considerably {\em faster} calculation (halving the time step more than doubles the CPU time required). 
\begin{table}[h]%% placement specifier
%% Use tabular environment to tag the tabular data.
%% https://en.wikibooks.org/wiki/LaTeX/Tables#The_tabular_environment
\centering%% For centre alignment of tabular.
\begin{tabular} {|c|c|c|c|c| }
\hline%% Table column specifiers
%% Tabular cells are separated by &
$\tau$ ($\tau_c$) & K. E. (K) & P. E. (K) & $\rho_S$ & $n_0$ \\ \hline
$2.0$ &13.823(7) &$-21.3616(8)$ &$1.03(4)$ & $0.070(1)$ \\
$0.25$ &$14.133(11)$ &$-21.3623(8)$ &$0.99(4)$ &$0.070(1)$\\
\hline 
\end{tabular}
%% Use \caption command for table caption and label.
\caption{Kinetic (K. E.) and potential (P. E.) energy per particle (in K), superfluid ($\rho_S$) and condensate ($n_0$) fraction  computed at the thermodynamic conditions of Fig. \ref{kextrap} for two values of the time step $\tau$ (expressed in units of $\tau_c=3.125\times 10^{-3}$ K$^{-1}$). Statistical errors, in parentheses, are on the last digit(s).}\label{table}
\end{table} 
\begin{figure}[h]
\centering
\includegraphics[width=0.7\linewidth]{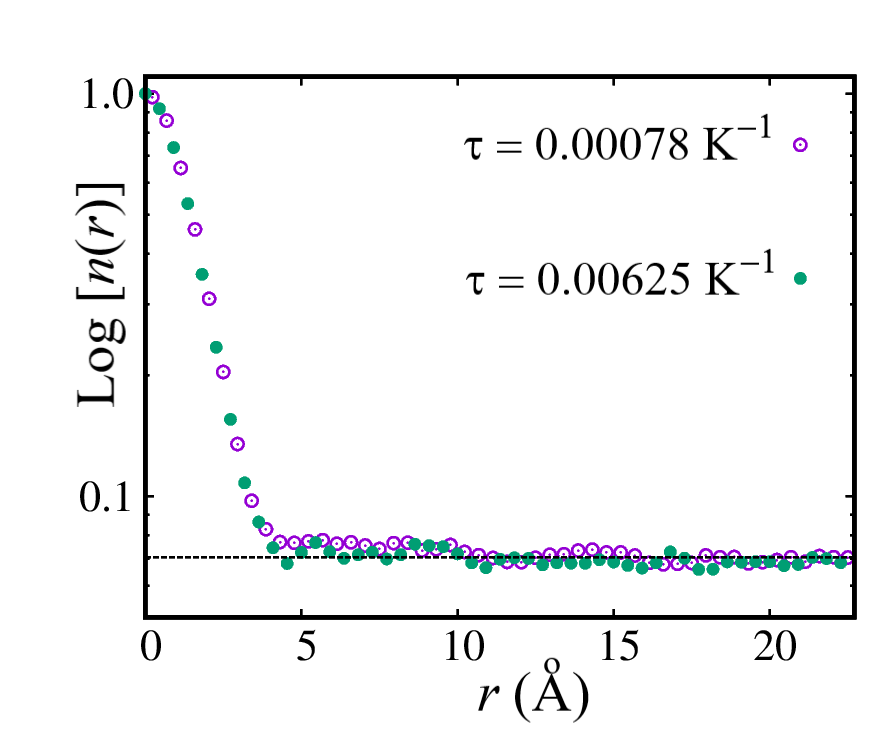}
\caption{One-body density matrix in superfluid $^4$He at the same thermodynamic conditions of Fig. \ref{kextrap}, with the two different values of the time step of Table \ref{table}.  Statistical errors are of the order of the size of the symbols.
The dashed line corresponds to the value 0.0705, which is consistent with the averages of the two sets of data. \label{nofrextrap}}
\end{figure} 
\\ \indent
To illustrate this point, Table \ref{table} reports the estimates for different physical quantities, computed by simulation at the thermodynamic conditions of Fig. \ref{kextrap}, for two different values of the time step (expressed in units of $\tau_c$). As one can see, while the kinetic energy per $^4$He atom computed with the shorter time step is within statistical errors consistent with the extrapolated $\tau=0$ value, that obtained with the longest time step utilized in this study  is merely $\sim 2\%$ higher. Whether such an error can be regarded as acceptable or not clearly depends on the scope of one's individual, specific investigation.
However, the results for the potential energy per particle, superfluid ($\rho_S$) and condensate  ($n_0$) fraction are all indistinguishable, within statistical uncertainties.
\\ \indent
This holds true not only for averaged quantities, but for correlation functions as well. Fig. \ref{nofrextrap} shows the one-particle density matrix $n(r)$ computed with the same two time steps; the results are virtually indistinguishable (note the logarithmic scale on the vertical axis). 
Summarizing, the choice of optimal time step is one that should be made based on the physical quantities one is interested in. Optimizing $\tau$ based on the energy (or other quantities that are related to it, such as the pressure) when one is ultimately interested in other quantities, is likely to result in unnecessarily long simulations, most of the time. 
\\ \indent
One last comment is worth making to conclude this discussion. It was  asserted \cite{Ceperley1995}  and for a long time believed, that the very {\em feasibility} of QMC simulations of strongly interacting many-body systems in continuous 
space required a sufficiently accurate short-time propagator, in order to keep the number of imaginary time slices down to a minimum (e.g., 40). Too large a number thereof may render path sampling inefficient, in turn leading to poor convergence of the estimates, especially at low temperature. The so-called pair product approximation (PPA), based on the exact numerical solution of the Bloch's equation for two particles, was proposed as the  optimal choice of short-time propagator. It was contented to deliver converged results for superfluid $^4$He  with a time step as large as 0.025 K$^{-1}$ (a statement  never convincingly, {\em quantitatively} demonstrated).
\\ \indent
However, subsequent experience has shown that reliable convergence can be achieved with relatively simple propagators such as the one used in this work, as shown by the results presented here. It seems fair to state that the need and effectiveness of the PPA, which is quite cumbersome to implement and of limited applicability, have been {\em vastly} overstated. Convergence problems affecting early implementations of continuous-space QMC were mainly due to inefficient path sampling, rather than to a large number of time slices; the WA allows one to overcome them without the use of {\em ad hoc}, abstruse devices.

\section{Results}\label{res}
\subsection{Energetics}\label{ene}
\begin{table}[h]%% placement specifier
%% Use tabular environment to tag the tabular data.
%% https://en.wikibooks.org/wiki/LaTeX/Tables#The_tabular_environment
\centering%% For centre alignment of tabular.
\begin{tabular} {|c|c|c|c|c| }
\hline%% Table column specifiers
%% Tabular cells are s14.1247  0.006563214.1247  0.006563214.1247  0.006563214.1247  0.0065632eparated by &
$T$ (K) & $n$ (\AA$^{-3}$) & K. E. (K) &  E (K) & K. E. (expt) \\ \hline
0.25 &0.021837 &14.126(2) &$-7.233(2)$ &  \\
%0.5 & 0.021837 &$14.116(4)$ & $-7.243(4)$ &  \\ 
$1.0$ & 0.021834 &$14.125(7)$ &$-7.233(7)$ &$14.3(3)$\\
& & 14.201(22)$^\star$ &$-7.386(22)^\star$ &  \\ 
1.4 & 0.021837 &14.222(10) &$-7.149(10)$ & $14.4(5)$ \\ 
1.65 &0.021851 & 14.392(13) &$-7.017(13)$ &$14.6(6)$ \\ 
$2.0$ & 0.021910 & $15.080(4)$ & $-6.470(4)$  &14.9(6)  \\
\hline 
\end{tabular}
%% Use \caption command for table caption and label.
\caption{Kinetic (K. E.) and total (E) energy per particle (in K) computed by QMC at various temperatures, at saturated vapor pressure. The values of the density $n$ are taken from Table 1.2 of Ref. \cite{Barenghi1998}. Statistical errors, in parentheses, are on the last digit(s). All the results pertain to simulations of a system comprising $N=2048$ atoms, with a time step $\tau=\tau_c=3.125\times 10^{-3}$ K$^{-1}$. The result marked with a $\star$ was obtained using the Aziz II potential. Rightmost column shows experimental estimates of the kinetic energy per atom from an analysis of neutron scattering data, specifically Ref. \cite{Prisk2017}. Note that the experimental entry quoted at $T=1$ K refers to measurements performed at 1.09 K.}\label{bigtable}
\end{table} 
A set of representative energy results of our study is offered in Table \ref{bigtable}. The values of density utilized at the different temperatures are obtained from Ref. \cite{Barenghi1998}. All of the theoretically computed values in Table \ref{bigtable} are the results of simulations with $N=2048$ $^4$He atoms, interacting via the pairwise Aziz I potential, except for those featuring a $\star$, to indicate that the Aziz II was used in that calculation. The time step utilized in all calculations is $\tau=\tau_c$ \cite{makesure}.
We find that the (kinetic) energy per particle remains unchanged, within our statistical uncertainties, if the temperature is lowered below 1 K; thus, we may conclude that the values at $T=1$ K provide reliable ground state estimates. On taking the estimate at $T=0.25$ K as representative of the ground state, we obtain an energy per particle  of $-7.233(2)$ K, which is significantly lower ($\sim$ 0.11 K) than essentially all ground state QMC estimates published over the past few decades, based on the same pair potential \cite{Prisk2017,Kalos1981,Boronat1994}. Such a discrepancy cannot be accounted for by slight differences in density between the various calculations. 
\\ 
\indent
The size of the system considered here is {8 to 32 times} greater than that on which previous published QMC estimates were obtained. However, this by itself does {\em not} quantitatively account for the difference in energy between this work and previous studies. For example, a simulation carried out an a system of $N=128$ atoms
{\color{black} yields} 
%\sout{only slightly different} 
estimates for kinetic and 
%\sout{potential} 
{\color{black} total} energy {\color {black} per atom of 14.114(11) K and $-7.233(11)$ K, respectively, i.e., virtually {\em identical} to } 
%\sout{with respect to} 
those quoted in Table \ref{bigtable} for $N=2048$ atoms.  
\\ \indent
The following remarks can be made:
\begin{enumerate}
    \item{The $\sim 0.05$ K difference between this calculation and that of Ref. \cite{Prisk2017}, based on the same methodology utilized here, seems likely the result of the much longer simulation time allotted for the present work, resulting in better equilibration and a more reliable estimate of the statistical uncertainties, which were possibly underestimated in Ref. \cite{Prisk2017}.  }
    \item {Practically {\em all} ground state QMC energy estimates were obtained using methods (such as Diffusion Monte Carlo) based on the evolution in imaginary time of a finite population of random walkers. As extensively documented  in the literature \cite{Boninsegni2013,Moroni2014,Yu2024} this kind of methodology tends to yield energy estimates {\em above} those furnished by QMC techniques (either at zero or finite temperature), that do not make use of populations \cite{Boninsegni2012b,Boninsegni2001}. This is due to the combined effect of finite population size and the bias due to the trial wave function out of which the ground state is projected.}
\end{enumerate} 
Theoretically computed energy values can be compared with experimentally measured ones. The currently accepted  \cite{Brooks1998} energy per $^4$He atom in the superfluid phase at $T=0$ is $-7.17$ K, which seems in reasonable agreement with the estimate obtained here (the uncertainty in the experimentally determined value is unreported in the literature). 
{\color {black} It is worth reminding, however, that in principle there is no reason to expect that a microscopic model including only a two-body interaction should allow one to achieve experimental accuracy, for it is known that contributions arising from three-body terms are significant. We come back to this point below. }
The kinetic energy per $^4$He atoms can also be measured experimentally, specifically inferred from the momentum distribution, in turn accessible via neutron scattering \cite{Mook1974,Sears1982,Azuah1997,Glyde2000,Glyde2011,Beauvois2018}. In general, the comparison between theory and experiment is satisfactory, even though the relatively large experimental uncertainties set a limit to its significance.
\\ \indent
Also included in Table \ref{bigtable} are theoretical estimates obtained with the Aziz II pair potential. As mentioned in the introduction, this is a revised version of the original Aziz potential, aiming at relying less on phenomenological input and more on first principle quantum chemistry calculations.  It should be mentioned that the Aziz II is not the most recent in its class, i.e., there have been several successive refinements \cite{Przybytek2017}. The comparison is only provided to illustrate the main quantitative differences between the original Aziz potential and {\em ab initio} pair interactions. Typically, these potentials feature a deeper attractive well than the original Aziz potential, resulting in a slightly greater ($\sim 0.15$ K) binding energy per $^4$He at equilibrium (as shown in Table \ref{bigtable}).
\\ \indent 
In general, there is almost no noticeable difference between the results yielded by the various {\em ab initio} pair potentials and the (phenomenological) Aziz I potential, with the notable exception of the energy and the pressure, for which the Aziz I potential affords better agreement with experiment. All these potentials, however, yield lower values of the pressure than experimentally measured. For example, at $T=1$ K the result of the simulation with the Aziz I potential is $-0.34(3)$ bars, compared to the experimental $1.56 \times 10^{-4}$ bars \cite{Barenghi1998}. \\ \indent 
It should be noticed that the results for the total energy per atom and the pressure, as yielded by {\em ab initio} potentials, can be corrected and brought in closer agreement with experiment \cite{Moroni2000} by explicitly including three-body contributions to the potential energy to the Hamiltonian (\ref{u}). The same correction cannot be easily applied to the Aziz I potential, which in part already incorporates three-body contributions in an effective way.  Three-body terms have little or no effect on structural and superfluid properties, and even on the kinetic energy \cite{Chang2001,Boninsegni1994}, which is mainly sensitive to the repulsive part of the two-body interaction. 
\subsection{Structure}\label{str}

\begin{figure}[t]
\centering
\includegraphics[width=1.0\linewidth] {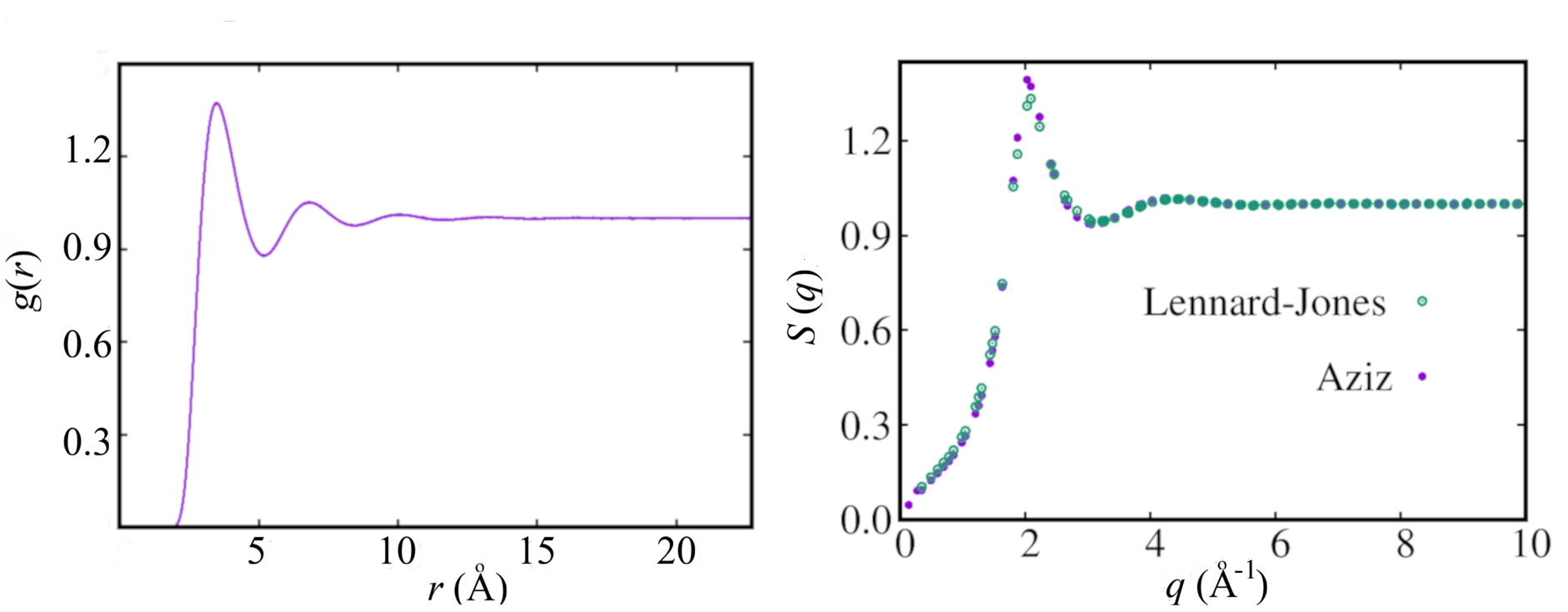}
\caption{Theoretically computed pair correlation function (left) and static structure factor (right) for superfluid $^4$He at temperature $T=1$ K, at saturated vapor pressure. The pair correlation function was computed using the Aziz I pair potential, whereas static structure factor was computed using both the Aziz I and Lennard-Jones potentials.
 \label{structure}}
\end{figure} 

The structure of a quantum fluid is generally quantitatively characterized by spatial correlation functions, chiefly the pair correlation function $g(r)$, and the (related) static structure factor $S(q)$, the latter accessible experimentally, through $x$-ray \cite{Robkoff1982} and  neutron \cite {Svensson1980} scattering. 
Theoretical results for these quantities, computed at temperature $T=1$ K at saturated vapor pressure are shown in Fig. \ref{structure}. The results presented here are not significantly different from those of previous QMC calculations on smaller system sizes, and therefore we refer interested readers to the discussion in, e.g., Ref. \cite {Ceperley1995}. As mentioned above, no noticeable difference exists between the results obtained with the Aziz I potential, shown in Fig. \ref{structure}, and those yielded by the Aziz II potential.
\\ \indent
Two specific aspects are worth pointing out, however. First, the pair correlation function is seen to converge to its asymptotic value (1) well within half of the size of the simulation cell (left part of Fig. \ref{structure}. This is of course one of the benefits of simulating a system comprising few thousand atoms, and gives confidence on the numerical accuracy of the approximation (i.e., setting $g(r)=1$) involved in estimating the contribution to the potential energy and the pressure arising from pairs of particles at distances $r > (L/2)$. 
\\ \indent
Regarding the static structure factor, the remarkable agreement between the result of a Path Integral Monte Carlo simulation and experimental measurements has been noted before, sometimes in almost celebratory tone \cite{Ceperley1995}. While this certainly provides validation of the microscopic model utilized in the calculation, chiefly the interatomic pair potential, its significance should not be overstated. As it turns out, integrated quantities such as the static structure factor display little sensitivity to the details of the pair potential. This is illustrated in the right part of Fig. \ref{structure}, in which the results obtained with the Aziz I pair potential are compared to those given by the much cruder Lennard-Jones interaction, which does not afford the same quantitative precision of the Aziz potential \cite{Kalos1981}. Except for a tiny difference in the height of the main peak, the results obtained for this quantity with the two potentials are virtually indistinguishable.
\subsection{Superfluidity and Bose Condensation}\label{sbc}
In Ref. \cite{Boninsegni2006b} the superfluid fraction was computed and its behavior quantitatively investigated. Upon performing accurate finite-size scaling analysis of results obtained for systems of different size, an estimate for the superfluid transition temperature $T_c$ was provided. 
Table \ref{rhontable} shows results obtained in this work (i.e., independently of the calculation of Ref. \cite{Boninsegni2006b}) for the $^4$He superfluid fraction $\rho_S(T)$ computed at various temperatures, at saturated vapor pressure. As mentioned in the abstract and in the introduction, the results  shown are for a system comprising $N=2048$ atoms.
\begin{table}[h]%% placement specifier
%% Use tabul ar environment to tag the tabular data.
%% https://en.wikibooks.org/wiki/LaTeX/Tables#The_tabular_environment
\centering%% For centre alignment of tabular.
\begin{tabular} {|c|c|c|c|c| }
\hline%% Table column specifiers
%% Tabular cells are separated by &
$T$ (K) & $\rho_S$ & $\rho_S$ (expt) & $n_0$ & $n_0$  (expt)\\ \hline
0.25 &0.99(2) &1.000 &0.071(1) & \\
0.5 &0.97(2) &1.000  &0.072(1) & \\
1.0 & 0.99(1) & 0.993 &0.070(1) &0.072(2) \\
1.4 &0.91(2) &0.926 & 0.068(1)& 0.072(4)\\
1.65 &0.79(2) &0.807 &0.062(1) &0.051(10) \\
2.0 &$0.45(2)$ &0.447 &0.039(1) &0.040(3) \\
\hline 
\end{tabular}
%% Use \caption command for table caption and label.
\caption{Theoretical (this work) and experimental estimates of the superfluid ($\rho_S$) and condensate fraction ($n_0$) in liquid $^4$He at various temperatures, at saturated vapor pressure. Experimental results for $\rho_S$ are taken from Ref. \cite{Barenghi1998}, those for $n_0$ are obtained as averages of values listed in Ref. \cite {Prisk2017}. }\label{rhontable}
\end{table} 

The superfluid fraction can be measured experimentally with a precision that is not presently attainable by QMC. Still, the agreement between theory and experiment is excellent for $T\lesssim$ 2 K; above this temperature, as the system approaches the transition, finite-size effects become important as the transition itself is ``smeared'', i.e., one obtains finite values of $\rho_S$ for $T > T_c$. In order to obtain an estimate of $T_c$ extrapolation of the results obtained on various systems of different sizes is required; as mentioned above, this was carried out in Ref. \cite{Boninsegni2006b}, and will not be repeated here.
Instead, we shall focus here on the behavior with temperature of the condensate fraction $n_0(T)$, which has not been quantitatively examined in previous QMC work, even though $n_0$ is in many respect even more fundamental a quantity, when it comes to characterizing the symmetry breaking character than the superfluid phase \cite{Babaev2015}. 
\begin{figure}[h]
\centering
\includegraphics[width=0.7\linewidth]{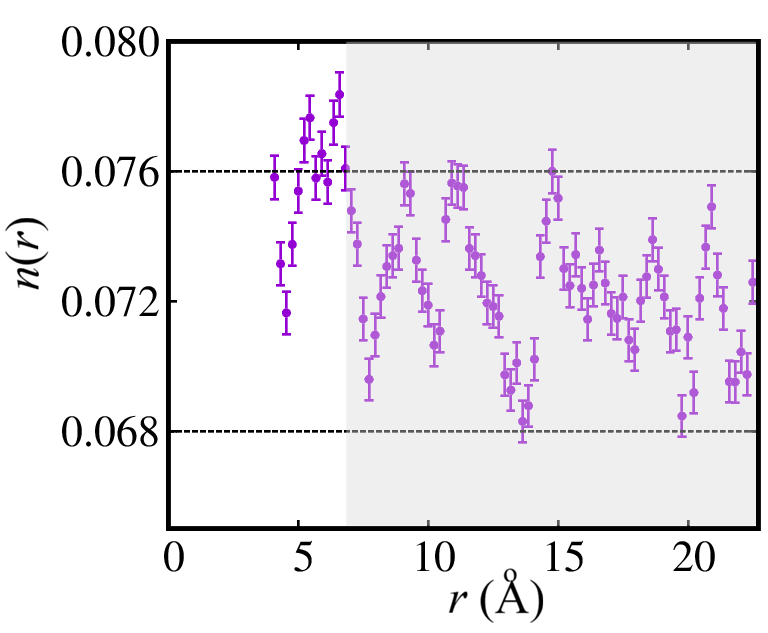}
\caption{One-body density matrix $n(r)$ computed by QMC at temperature $T=0.5$ K, at saturated vapor pressure, on a system comprising $N=2048$ atoms. The value of the condensate fraction is obtained by averaging over the values in the shaded area.
 \label{asy}}
\end{figure}
\\ \indent
In a 3D superfluid, the one-body density matrix $n(r)$ plateaus at large distances to a constant value $n_0$, which is the condensate fraction. Such a behavior is clearly observed in the results shown in Fig. \ref{nofrextrap}; indeed, on a logarithmic scale $n(r)$ appears constant, within statistical errors, for $r \gtrsim 5$ \AA. Closer inspection shows that $n(r)$ oscillates around the asymptotic value, as shown in Fig. \ref{asy}. The way in which $n(r)$ approaches its asymptotic value at large distances is described by Bogoliubov theory, which makes it possible to estimate the value $n_0^0$ of the condensate fraction at $T=0$ by fitting the {\em asymptotic} behavior at long distance of the $n(r)$ computed at a {\em single}, {sufficiently low temperature} to an expression containing $n_0^0$ as an adjustable parameter; this is the procedure adopted in Ref.  \cite{Boninsegni2006b} (Eq. 6.1 therein).
\\ \indent
In this work, on the other hand, we computed $n(r)$ for several temperature, and for each temperature estimated $n_0(T)$ by averaging $n(r)$ over a range of distances within which most of the values of $n(r)$ fall within a fairly well-defined interval (as in the shaded area of Fig. \ref{asy}, with all values falling between the dashed horizontal lines), indicating that $n(r)$ displays relatively small fluctuations around a  constant value. Specifically, we histogram the values of $n(r)$ used to compute the average, and quote as statistical error the semi-width of the $n(r)$ interval containing approximately 2/3 thereof.
\\ \indent
As the temperature is increased from $T=0$ toward $T_c$, such a plateauing of $n(r)$ occurs at greater and greater distance, eventually exceeding the side of the simulation cell, i.e., not being observable at all. 
Clearly, therefore, reliably investigating the behavior of $n_0(T)$ near $T_c$ requires that one be able to carry out simulations on systems of sufficiently large size, or, equivalently phrased, any given system sizes sets a limit on how closely one can approach $T_c$ and still obtain a reliable estimate of $n_0 (T)$.
\begin{figure}[h]
\centering
\includegraphics[width=0.7\linewidth]{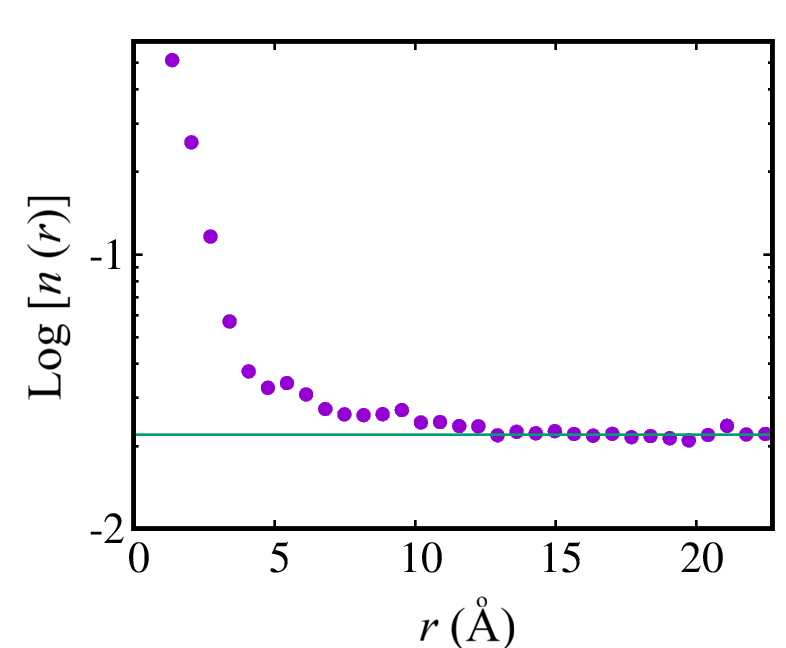}
\caption{One-body density matrix $n(r)$ computed by QMC at temperature $T=2.12$ K, at saturated vapor pressure, on a system comprising $N=2048$ atoms. Solid line corresponds to the value 0.022. Error bars are smaller than the size of the symbols.
 \label{highest}}
\end{figure} 
\\ \indent
Fig. \ref{highest} shows the one-body density matrix for the highest temperature considered in this work for which $n_0$ can be determined with reasonable confidence (see results reported in Table \ref{rhontable}), again by averaging results obtained for  distances $\gtrsim 10$ \AA. This temperature is within $\sim 3\%$ of the experimentally measured $T_c$ \cite{Barenghi1998}. The behavior of the $n(r)$ drastically changes at higher temperature; at the immediately higher temperature considered in this work, namely $T=2.18$ K, no clear plateau can be identified and within error bars the results are consistent with an exponential decay \cite{notesize}. This is, of course, consistent with experiment; however (perhaps more likely), it may also be due to the fact that the simulation cell is too small to observe a plateau, at this particular temperature. That is, $n_0$ may well be {\em finite} at this particular temperature, but this observation may require that one simulate a system of much larger size than just 2048 atoms. After all, the microscopic model utilized (Eq. \ref{u}) is not necessarily expected to reproduce all experimental results to this level of accuracy.

\begin{figure}[h]
\centering
\includegraphics[width=0.7\linewidth]{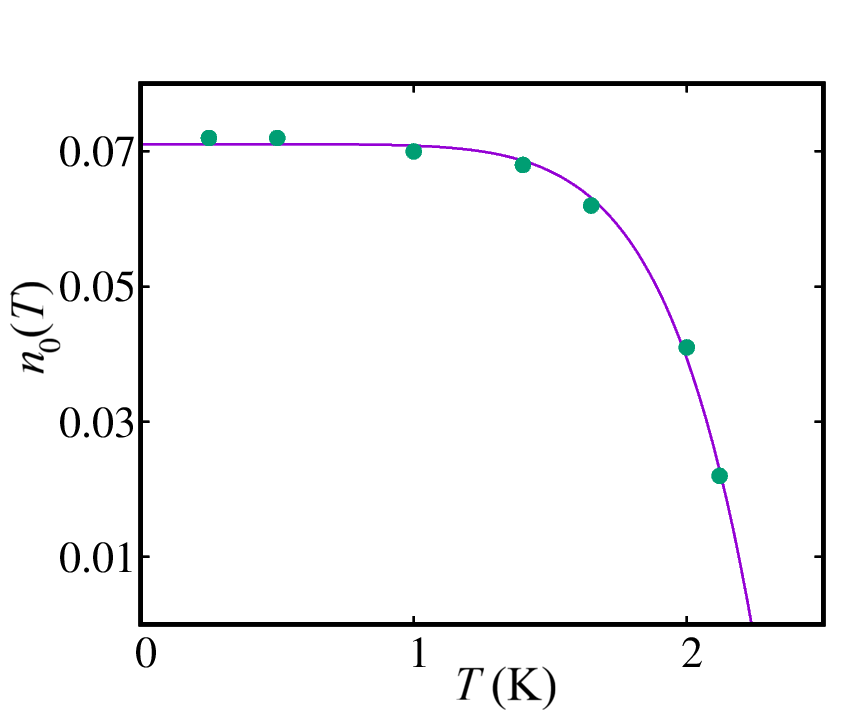}
\caption{Condensate fraction $n_0(T)$ computed by QMC at various temperatures, at saturated vapor pressure. Statistical errors are smaller than the size of the symbols. Solid line is a fit to the data obtained as described in the text.
 \label{cfrac}}
\end{figure} 

Experimentally, the condensate fraction has  been inferred from the measured atomic momentum distribution \cite{Prisk2017,Sears1982,Azuah1997,Glyde2000,Glyde2011,Snow1992}. Table \ref{rhontable} shows a comparison of our estimates of $n_0(T)$ with a compilation of the most recent published measurements, from Ref. \cite{Prisk2017}. The agreement is quite satisfactory. Our results for $n_0(T)$ are also shown in  Fig. \ref{cfrac}. The first, clear observation is that, within the statistical errors of our calculations $n_0$ does not change significantly below $T\sim 1.4$ K, consistently with experimental observation. At low temperature, it is seen to remain close to $\sim 7\%$, i.e., a value significantly {\em lower} than that ($\sim 8\%$) estimated in Ref. \cite{Boninsegni2006b} as the ground state value $n_0^0$, based on the study of the oscillatory behavior at long distances of $n(r)$, computed for a single value of $T$ (1 K).
\\ \indent
Although the calculation of Ref. \cite{Boninsegni2006b} and the one carried out in this work are based on the same methodology, they differ in some important aspects. The most obvious is that the current work benefits from more advanced computing equipment; in particular, the resources and time frame allocated to the work published 20 years ago did not allow for the same detailed study of the condensate fraction as a function of temperature presented here (at least not by simulating a system of the size considered here). Thus, the value of $n_0^0$ was inferred as explained above, i.e., by fitting the asymptotic behavior at long distance of the $n(r)$ computed at a {\em single temperature}. Specifically, the fit was performed for a system comprising $N=1024$ atoms at temperature $T=1$ K, assumed large and low enough that the Bogoliubov theory should hold. Based on the results obtained in this work, however, it seems that either or both assumptions may not have been entirely valid.
\\ \indent
The question immediately arises of whether the data obtained here on a larger system size and lower temperature can be utilized to arrive at a more reliable estimation of $n_0^0$, and possibly of the superfluid transition temperature $T_c$, at which {\em both} $n_0$ and $\rho_S$ vanish. One approach which has been used, based on experimental measurements of $n_0(T)$, has consisted of fitting the data  with the expression \cite{Sears1982,Glyde2013}
\begin{equation}
n_0(T)=n_0^0 \ \biggl [1-\biggl ( \frac{T}{T_c}\biggr)^\gamma\biggr], 
\label{empi}
\end{equation}
where $n_0^0, T_c$ and $\gamma$ are fitting parameters. The motivation for making use of such an expression is that it is exact for the non-interacting Bose gas, in which case $\gamma=3$; to our knowledge, it is {\em ad hoc}, i.e., it has no theoretical foundation for a strongly interacting Bose fluid such as $^4$He. However, it provides a good fit to our data, as shown in Fig. \ref{cfrac}; the values of the fitting parameters are $n_0^0=0.0715\pm0.0004$,  $T_c=2.25\pm0.03$ and $\gamma=6.9\pm0.7$. 
\\ \indent
The most precisely and reliably determined of these parameters is $n_0^0$, which is the best estimate we can offer of the condensate fraction in the ground state of superfluid $^4$He, based on the results are presented here.  Because $n_0(T)$ is essentially constant below 1.4 K, it can be confidently assumed that the value arising from the fit is robust, i.e., relatively insensitive to the details of the fitting form utilized (the same cannot be said of the other two fitting parameters).
As mentioned, this estimate for $n_0^0$ is 12.5\% below the indirect one of Ref. \cite{Boninsegni2006b}, but is entirely consistent with the most recent experimental estimates. It is also in agreement, taking into account combined statistical and systematic (finite size) uncertainties, with the QMC estimates at finite temperature of Ref. \cite{Prisk2017} and at zero temperature of Ref. \cite{Moroni2004}, both the results  of simulations of (an order of magnitude) smaller systems than that considered here.
\\ \indent
Reasonably well determined is also the transition temperature $T_c$, whose value is above the experimentally known one as it should be expected, given the finite size of the system studied. This value is understandably quite sensitive to values of $n_0(T)$ in the vicinity of $T_c$, where finite-size effects are most important. 
The parameter $\gamma$, which unlike $n_0^0$ and $T_c$ is not connected to any experimental quantity, is the least well determined, as it is quite sensitive to the behavior of $n_0(T)$ in the entire temperature range. Nevertheless, its value is in the same ballpark as in previous work, in which experimental estimates of $n_0(T)$ were used.
\\ \indent
It should be stressed that the proper way to arrive at a robust estimation of the superfluid transition temperature is by performing rigorous finite size scaling analysis of results for the superfluid fraction (as illustrated in detail in Ref. \cite{Boninsegni2006b}), for which a reliable theory exists based on the winding number and its topological properties. Given that such a finite-size scaling analysis is often fairly onerous and/or impractical from the computational standpoint, the purpose of this study is to provide some quantitative measure of ``how far'' from the thermodynamic limit can be considered the results of a simulation of a system of size such as that utilized here. 
Altogether, it seems fair to state that they provide a reasonably accurate quantitative estimate of $T_c$, within the obvious limitation of using the empirical fitting form (\ref{empi}). 
\section{Conclusions}\label{concl}
In this paper we present results of extensive numerical studies based on the continuous-space WA, of the low-temperature properties of the superfluid phase of $^4$He at saturated vapor pressure.\\ \indent The purpose of this work is twofold; on the one hand, we provide accurate numerical results for energetics, structure and the superfluid properties of the system, that have not been published so far, at least not as comprehensively and on a system of the size considered in this work. Without exception, all available published results were obtained on systems too small (less than 100 atoms) to yield results reliably representative of the thermodynamic limit, or using ground state computational methodologies (e.g., DMC) long advertised as ``exact'' when in practice they are affected by now well-documented systematic errors associated to the finite population of random walkers, as well as by bias due to the use of a trial wave function. Indeed, just like in a previous study of the solid phase of $^4$He \cite{Boninsegni2023b}, significant numerical differences are reported between the energy estimates obtained here and those published decades ago, based on, e.g., GFMC, and never revisited and/or revised. We also provide a revised estimate of the condensate fraction at $T=0$, namely 0.0715$\pm 0.0004$, which is in agreement with the most recent experimental data, albeit substantially lower than the most current existing estimates (it is only consistent with that of Ref. \cite{Moroni2004}, which is however affected by significantly greater uncertainties). We believe that the difference with previous calculations arises from the greater system size utilized in this work, as well as the fact that calculations at considerably lower temperature were carried out.\\ \indent
The second goal is to assess quantitatively the present state of the art of QMC simulations for Bose systems, and hopefully set the standard for what should at this point be regarded as the {\em minimum} system size and typical temperature range for future comparable studies.  We offer results which will hopefully prove useful benchmarks for researchers developing new theoretical methods for quantum many-body systems.
We have focused on quantities for which in principle exact numerical results can be provided, only affected by relatively small systematic and statistical errors. We have left out of this discussion quantities such as the dynamic structure factor, which in principle could be inferred by performing an inversion of QMC-computed imaginary-time correlations functions, but it seems fair to state that no well-defined procedure has been proposed yet to overcome the inherent ill-posedness of the problem. Interested readers are referred to Refs. \cite{Boninsegni1996,Galli2010,Kora2018}.
\\ \indent
This work was supported by the Natural Science and Engineering Research Council of Canada under grant RGPIN 2024-05664.

\bibliographystyle{unsrt}
\bibliography{refs}

@article{Pollock1987,
  title = {Path-integral computation of superfluid densities},
  author = {Pollock, E. L. and Ceperley, D. M.},
  journal = {Phys. Rev. B},
  volume = {36},
  issue = {16},
  pages = {8343--8352},
  numpages = {0},
  year = {1987},
  publisher = {American Physical Society},
  doi = {10.1103/PhysRevB.36.8343},
  url = {https://link.aps.org/doi/10.1103/PhysRevB.36.8343}
}

@article{Moroni2004,
   author = {Moroni, S. and Boninsegni, M.},
   title = {{Condensate Fraction in Liquid} $^4${He}},
   journal = {J. Low Temp. Phys.},
   volume = {136},
   number = {3/4},
   pages = {129–137},
   DOI = {10.1023/B:JOLT.0000038518.10132.30},
   year = {2004},
   type = {Journal Article}
}

@article{Boninsegni1997,
   author = {Boninsegni, M. and Moroni, S.},
   title = {{Microscopic Calculation of Superfluidity and Kinetic Energies in Isotopic Liquid Helium Mixtures}},
   journal = {Phys. Rev. Lett.},
   volume = {78},
   number = {9},
   pages = {1727–1730},
   DOI = {10.1103/PhysRevLett.78.1727},
   year = {1997},
   type = {Journal Article}
}

@book{Babaev2015,
    author = {Svistunov, B. V. and Babaev, 
    E. S. and Prokof'ev, N. V.},
    title = {Superfluid States of Matter},
    publisher = {CRC Press},
    doi = {10.1201/b18346},
    edition = {first},
    address = {Boca Raton},
    year = {2015}
}

@article{Ceperley1986,
   author = {Ceperley, D. M. and Pollock, E. L.},
   title = {{Path-integral Computation of the Low-Temperature Properties of Liquid $^4$He}},
   journal = {Phys. Rev. Lett.},
   volume = {56},
   number = {4},
   pages = {351–354},
   year = {1986},
   type = {Journal Article}
}

@article{Boninsegni2006,
  title = {{Worm Algorithm for Continuous-Space Path Integral Monte Carlo Simulations}},
  author = {Boninsegni, M. and Prokof'ev, N. and Svistunov, B.},
  journal = {Phys. Rev. Lett.},
  volume = {96},
  issue = {7},
  pages = {070601},
  numpages = {4},
  year = {2006},
  publisher = {American Physical Society},
  doi = {10.1103/PhysRevLett.96.070601},
  url = {https://link.aps.org/doi/10.1103/PhysRevLett.96.070601}
}

@article{Boninsegni2006b,
  title = {Worm algorithm and diagrammatic Monte Carlo: A new approach to continuous-space path integral {Monte Carlo} simulations},
  author = {Boninsegni, M. and Prokof'ev, N. V. and Svistunov, B. V.},
  journal = {Phys. Rev. E},
  volume = {74},
  issue = {3},
  pages = {036701},
  numpages = {16},
  year = {2006},
  publisher = {American Physical Society},
  doi = {10.1103/PhysRevE.74.036701}
}

@article{Saccani2011,
   author = {Saccani, S. and Moroni, S. and Boninsegni, M.},
   title = {Phase diagram of soft-core bosons in two dimensions},
   journal = {Phys. Rev. B},
   volume = {83},
   number = {9},
   pages = {92506},
   DOI = {10.1103/PhysRevB.83.092506},
   url = {http://link.aps.org/doi/10.1103/PhysRevB.83.092506},
   year = {2011},
   type = {Journal Article}
}

@article{Boninsegni2012c,
   author = {Boninsegni, M.},
   title = {Supersolid Phases of Cold Atom Assemblies},
   journal = {J. Low Temp. Phys.},
   volume = {168},
   number = {3-4},
   pages = {137-149},
   DOI = {10.1007/s10909-012-0571-1},
   year = {2012},
   type = {Journal Article}
}

@article{Kora2019,
   author = {Kora, Y. and Boninsegni, M.},
   title = {Patterned Supersolids in Dipolar Bose Systems},
   journal = {J. Low Temp. Phys.},
   volume = {197},
   number = {5-6},
   pages = {337–347},
   DOI = {10.1007/s10909-019-02229-z},
   url = {http://link.springer.com/10.1007/s10909-019-02229-z},
   year = {2019},
   type = {Journal Article}
}

@article{Filinov2010,
   author = {Filinov, A. V. and Prokof’ev, N. V. and Bonitz, M.},
   title = {{Berezinskii-Kosterlitz-Thouless Transition in Two-Dimensional Dipole Systems}},
   journal = {Phys. Rev. Lett.},
   volume = {105},
   number = {7},
   pages = {70401},
   url = {papers2://publication/doi/10.1103/PhysRevLett.105.070401},
   year = {2010},
   type = {Journal Article}
}

@article{Zhang2023,
   author = {Zhang, C. and Capogrosso-Sansone, B. and Boninsegni, M. and Prokof'ev, N. V. and Svistunov, B. V.},
   title = {{Superconducting Transition Temperature of the Bose One-Component Plasma}},
   journal = {Phys. Rev. Lett.},
   volume = {130},
   number = {23},
   pages = {236001},
   DOI = {10.1103/PhysRevLett.130.236001},
   year = {2023},
   type = {Journal Article}
}

@article{Boninsegni2026,
   author = {Boninsegni, M.},
   title = {{Bose one-component plasma in 2D: a Monte Carlo study}},
   journal = {Phys. Rev. B},
   volume = {113},
   pages = {094508},
   DOI = {10.1103/9rpx-wdjs},
   year = {2026},
   type = {Journal Article}
}

@inbook{Feynman1965,
  author    = {Feynman, R. P. and Hibbs, A. R.},
  title = {Quantum Mechanics and Path Integrals},
  publisher = {McGraw-Hill},
  year      = {1965},
  address   = {New York},
  chapter   = {10},
}

@article{Ceperley1995,
   author = {Ceperley, D. M.},
   title = {Path integrals in the theory of condensed helium},
   journal = {Rev. Mod. Phys.},
   volume = {67},
   number = {2},
   pages = {279–355},
   DOI = {10.1103/RevModPhys.67.279},
   year = {1995},
   type = {Journal Article}
}

@article{Barenghi1998,
   author = {Donnelly, R. J. and Barenghi, C. F.},
   title = {The Observed Properties of Liquid Helium at the Saturated Vapor Pressure},
   journal = {J. Phys. Chem. Ref. Data},
   volume = {27},
   number = {6},
   pages = {1217–1274},
   DOI = {10.1063/1.556028},
   year = {1998},
   type = {Journal Article}
}

@article{Feynman1953,
   author = {Feynman, R. P.},
   title = {Atomic Theory of Liquid Helium Near Absolute Zero},
   journal = {Phys. Rev.},
   volume = {91},
   number = {6},
   pages = {1301–1308},
   year = {1953},
   type = {Journal Article}
}

@book{Wilks1967,
  author    = {Wilks, J.},
  title     = {The Properties of Liquid and Solid Helium},
  publisher = {Clarendon Press},
  year      = {1967},
  address   = {Oxford}
}

@article{Aziz1979,
    author = {Aziz, R. A. and Nain, V. P. S. and Carley, J. S. and Taylor, W. L. and McConville, G. T.},
    title = "{An accurate intermolecular potential for helium}",
    journal = {The Journal of Chemical Physics},
    volume = {70},
    number = {9},
    pages = {4330-4342},
    year = {1979},
    doi = {10.1063/1.438007}
}

@article{Cencek2007,
   author = {Cencek, W. and Jeziorska, M. and Akin-Ojo, O. and Szalewicz, K.},
   title = {Three-Body Contribution to the Helium Interaction Potential},
   journal = {J. Phys. Chem. A},
   volume = {111},
   number = {44},
   pages = {11311-11319},
   ISSN = {1089-5639},
   DOI = {10.1021/jp072106n},
   url = {https://dx.doi.org/10.1021/jp072106n},
   year = {2007},
   type = {Journal Article}
}

@Article{Sese2020,
AUTHOR = {Sesé, L. M.},
TITLE = {Real Space Triplets in Quantum Condensed Matter: Numerical Experiments Using Path Integrals, Closures, and Hard Spheres},
JOURNAL = {Entropy},
VOLUME = {22},
YEAR = {2020},
NUMBER = {12},
ARTICLE-NUMBER = {1338},
URL = {https://www.mdpi.com/1099-4300/22/12/1338},
PubMedID = {33266522},
ISSN = {1099-4300},
DOI = {10.3390/e22121338}
}

@article{Moroni2000,
  title = {Equation of State of Solid ${}^{3}\mathrm{He}$},
  author = {Moroni, S. and Pederiva, F. and Fantoni, S. and Boninsegni, M.},
  journal = {Phys. Rev. Lett.},
  volume = {84},
  issue = {12},
  pages = {2650--2653},
  numpages = {0},
  year = {2000},
  publisher = {American Physical Society},
  doi = {10.1103/PhysRevLett.84.2650}
}

@article{Chang2001,
   author = {Chang, S. Y. and Boninsegni, M.},
   title = {Ab initio potentials and the equation of state of condensed helium at high pressure},
   journal = {J. Chem. Phys.},
   volume = {115},
   number = {6},
   pages = {2629–2633},
   DOI = {10.1063/1.1386657},
   year = {2001},
   type = {Journal Article}
}

@article{Kalos1974,
  title = {Helium at zero temperature with hard-sphere and other forces},
  author = {Kalos, M. H. and Levesque, D. and Verlet, L.},
  journal = {Phys. Rev. A},
  volume = {9},
  issue = {5},
  pages = {2178--2195},
  numpages = {0},
  year = {1974},
  publisher = {American Physical Society},
  doi = {10.1103/PhysRevA.9.2178}
}

@article{Boninsegni1994,
   author = {Boninsegni, M. and Pierleoni, C. and Ceperley, D. M.},
   title = {Isotopic shift of helium melting pressure: Path integral Monte Carlo study},
   journal = {Phys. Rev. Lett.},
   volume = {72},
   number = {12},
   pages = {1854–1857},
   DOI = {10.1103/PhysRevLett.72.1854},
   year = {1994},
   type = {Journal Article}
}

@article{Snow1992,
doi = {10.1209/0295-5075/19/5/010},
year = {1992},
volume = {19},
number = {5},
pages = {403},
author = {W. M. Snow and Y. Wang and P. E. Sokol},
title = {{Density and Temperature Dependence of the Condensate Fraction in Liquid} $^4${He}},
journal = {Europhys. Lett.}
}

@article{Robkoff1982,
  title = {Structure-factor measurements in $^{4}\mathrm{He}$ as a function of density},
  author = {Robkoff, H. N. and Hallock, R. B.},
  journal = {Phys. Rev. B},
  volume = {25},
  issue = {3},
  pages = {1572--1588},
  numpages = {0},
  year = {1982},
  publisher = {American Physical Society},
  doi = {10.1103/PhysRevB.25.1572}
}

@article{Svensson1980,
   author = {Svensson, E. C. and Sears, V. F. and Woods, A. D. B. and Martel, P.},
   title = {{Neutron-diffraction study of the static structure factor and pair correlations in liquid He-4}},
   journal = {Phys. Rev. B},
   volume = {21},
   number = {8},
   pages = {3638–3651},
   DOI = {10.1103/PhysRevB.21.3638},
   year = {1980},
   type = {Journal Article}
}

@article{Sears1982,
  title = {Neutron-Scattering Determination of the Momentum Distribution and the Condensate Fraction in Liquid $^{4}\mathrm{He}$},
  author = {Sears, V. F. and Svensson, E. C. and Martel, P. and Woods, A. D. B.},
  journal = {Phys. Rev. Lett.},
  volume = {49},
  issue = {4},
  pages = {279--282},
  numpages = {0},
  year = {1982},
  publisher = {American Physical Society},
  doi = {10.1103/PhysRevLett.49.279}
}

@article{Glyde2013,
    title = {{Bose-Einstein Condensation Measurements and Superflow in Condensed Helium}},
    author = {Glyde, H. R.},
    journal = {J. Low Temp. Phys.},
    year = {2013},
    volume = {172},
    pages = {364}
}

@article{Przybytek2017,
  title = {Pair Potential with Submillikelvin Uncertainties and Nonadiabatic Treatment of the Halo State of the Helium Dimer},
  author = {Przybytek, M. and Cencek, W. and Jeziorski, B. and Szalewicz, K.},
  journal = {Phys. Rev. Lett.},
  volume = {119},
  issue = {12},
  pages = {123401},
  numpages = {6},
  year = {2017},
  publisher = {American Physical Society},
  doi = {10.1103/PhysRevLett.119.123401}
}

@article{Kalos1981,
   author = {Kalos, M. H. and Lee, M. A. and Whitlock, P. A.},
   title = {Modern potentials and the properties of condensed Helium-four},
   journal = {Phys. Rev. B},
   volume = {24},
   number = {1},
   pages = {115–130},
   year = {1981},
   type = {Journal Article}
}

@article{Aziz1995,
  title = {Ab Initio Calculations for Helium: A Standard for Transport Property Measurements},
  author = {Aziz, R. A. and Janzen, A. R. and Moldover, M. R.},
  journal = {Phys. Rev. Lett.},
  volume = {74},
  issue = {9},
  pages = {1586--1589},
  numpages = {0},
  year = {1995},
  month = {Feb},
  publisher = {American Physical Society},
  doi = {10.1103/PhysRevLett.74.1586}
}

@article{Mezzacapo2006,
  title = {Superfluidity and Quantum Melting of {\em p}-{H}$_2$ Clusters},
  author = {Mezzacapo, F. and Boninsegni, M.},
  journal = {Phys. Rev. Lett.},
  volume = {97},
  issue = {4},
  pages = {045301},
  numpages = {4},
  year = {2006},
  publisher = {American Physical Society},
  doi = {10.1103/PhysRevLett.97.045301},
  url = {https://link.aps.org/doi/10.1103/PhysRevLett.97.045301}
}

@article{Mezzacapo2007,
  title = {Structure, superfluidity, and quantum melting of hydrogen clusters},
  author = {Mezzacapo, F. and Boninsegni, M.},
  journal = {Phys. Rev. A},
  volume = {75},
  issue = {3},
  pages = {033201},
  numpages = {10},
  year = {2007},
  month = {Mar},
  publisher = {American Physical Society},
  doi = {10.1103/PhysRevA.75.033201},
  url = {https://link.aps.org/doi/10.1103/PhysRevA.75.033201}
}

@article{Boninsegni2005,
   author = {Boninsegni, M.},
   title = {Permutation Sampling in {Path Integral Monte Carlo}},
   journal = {J. Low Temp. Phys.},
   volume = {141},
   number = {1-2},
   pages = {27-46},
   DOI = {10.1007/s10909-005-7513-0},
   url = {http://link.springer.com/10.1007/s10909-005-7513-0},
   year = {2005},
   type = {Journal Article}
}

@note{comment,
    note = {Usually the total, rather than the kinetic energy is extrapolated, but the procedures are equivalent because the kinetic energy is by far the most important source of time step error, as shown in Table \ref{table}.}
}

@note{makesure,
    note = {Extrapolation to the $\tau=0$ limit can be carried out for the total energy, just as illustrated in Fig. \ref{kextrap} for the kinetic energy. In practice, the results for the total energy obtained with $\tau\le\tau_c$ are indistinguishable within statistical uncertainties.}
}

@article{Moroni2014,
   author = {Moroni, S. and Boninsegni, M.},
   title = {Coexistence, Interfacial Energy, and the Fate of Microemulsions of 2D Dipolar Bosons},
   journal = {Phys. Rev. Lett.},
   volume = {113},
   number = {24},
   pages = {240407–240407},
   DOI = {10.1103/PhysRevLett.113.240407},
   year = {2014},
   type = {Journal Article}
}

@article{Brooks1998,
  author    = {Brooks, J. S. and Donnelly, R. J.},
  title     = {The Observed Properties of Liquid Helium at the Saturated Vapor Pressure},
  journal   = {J. Phys. Chem. Ref. Data},
  volume    = {27},
  number    = {6},
  pages     = {1217--1274},
  year      = {1998},
  doi       = {10.1063/1.556028}
}

@article{Glyde2000,
   author = {Glyde, H. R. and Azuah, R. T. and Stirling, W. G.},
   title = {Condensate, momentum distribution and final-state effects in liquid Helium-four},
   journal = {Phys. Rev. B},
   volume = {62},
   number = {21},
   pages = {14337–14349},
   year = {2000},
   type = {Journal Article}
}

@article{Glyde2011,
  title = {{Atomic momentum distribution and Bose-Einstein condensation in liquid ${}^{4}$He under pressure}},
  author = {Glyde, H. R. and Diallo, S. O. and Azuah, R. T. and Kirichek, O. and Taylor, J. W.},
  journal = {Phys. Rev. B},
  volume = {84},
  issue = {18},
  pages = {184506},
  numpages = {14},
  year = {2011},
  publisher = {American Physical Society},
  doi = {10.1103/PhysRevB.84.184506}
}

@article{Beauvois2018,
  title = {Microscopic dynamics of superfluid $^{4}\mathrm{He}$: A comprehensive study by inelastic neutron scattering},
  author = {Beauvois, K. and Dawidowski, J. and F\aa{}k, B. and Godfrin, H. and Krotscheck, E. and Ollivier, J. and Sultan, A.},
  journal = {Phys. Rev. B},
  volume = {97},
  issue = {18},
  pages = {184520},
  numpages = {14},
  year = {2018},
  publisher = {American Physical Society},
  doi = {10.1103/PhysRevB.97.184520}
}

@article{Prisk2017,
   author = {Prisk, T. R. and Bryan, M. S. and Sokol, P. E. and Granroth, G. E. and Moroni, S. and Boninsegni, M.},
   title = {{The Momentum Distribution of Liquid He-4}},
   journal = {J. Low Temp. Phys.},
   volume = {189},
   number = {3-4},
   pages = {158–184},
   DOI = {10.1007/s10909-017-1798-7},
   url = {http://link.springer.com/10.1007/s10909-017-1798-7},
   year = {2017},
   type = {Journal Article}
}

@article{Boronat1994,
   author = {Boronat, J. and Casulleras, J.},
   title = {{Monte Carlo analysis of an interatomic potential for He}},
   journal = {Phys. Rev. B},
   volume = {49},
   number = {13},
   pages = {8920–8930},
   url = {papers2://publication/uuid/0AEF01D4-C0A5-4D4D-801E-5CD02E4BD726},
   year = {1994},
   type = {Journal Article}
}

@article{Mook1974,
  title = {Neutron-Scattering Study of the Momentum Distribution of $^{4}\mathrm{He}$},
  author = {Mook, H. A.},
  journal = {Phys. Rev. Lett.},
  volume = {32},
  issue = {21},
  pages = {1167--1170},
  numpages = {0},
  year = {1974},
  publisher = {American Physical Society},
  doi = {10.1103/PhysRevLett.32.1167}
}

@article{Azuah1997,
   author = {Azuah, R. T. and Stirling, W. R. and Glyde, H. R. and Boninsegni, M. and Sokol, P. E. and Bennington, S. M.},
   title = {Condensate and final-state effects in superfluid $^4${He}},
   journal = {Phys. Rev. B},
   volume = {56},
   number = {22},
   pages = {14620–14630},
   DOI = {10.1103/PhysRevB.56.14620},
   year = {1997},
   type = {Journal Article}
}

@article{Boninsegni2013,
   author = {Boninsegni, M.},
   title = {Ground State Phase Diagram of Parahydrogen in One Dimension},
   journal = {Phys. Rev. Lett.},
   volume = {111},
   number = {23},
   pages = {235303},
   DOI = {10.1103/PhysRevLett.111.235303},
   url = {http://link.aps.org/doi/10.1103/PhysRevLett.111.235303},
   year = {2013},
   type = {Journal Article}
}

@article{Galli2010,
   author = {Vitali, E. and Rossi, M. and Reatto, L. and Galli, D. E.},
   title = {{Ab initio low-energy dynamics of superfluid and solid $^4$He}},
   journal = {Phys. Rev. B},
   volume = {82},
   number = {17},
   pages = {174510–174510},
   DOI = {10.1103/PhysRevB.82.174510},
   year = {2010},
   type = {Journal Article}
}

@article{Kora2018,
   author = {Kora, Y. and Boninsegni, M.},
   title = {{Dynamic structure factor of superfluid He-4 from quantum Monte Carlo: Maximum entropy revisited}},
   journal = {Phys. Rev. B},
   volume = {98},
   number = {13},
   pages = {134509–134509},
   DOI = {10.1103/PhysRevB.98.134509},
   year = {2018},
   type = {Journal Article}
}

@article{Boninsegni1996,
   author = {Boninsegni, M. and Ceperley, D. M.},
   title = {{Density fluctuations in liquid $^4$He. Path integrals and maximum entropy}},
   journal = {J. Low Temp. Phys.},
   volume = {104},
   number = {5-6},
   pages = {339–357},
   DOI = {10.1007/BF00751861},
   year = {1996},
   type = {Journal Article}
}

@note{notesize,
    note = {Generally speaking, in the $T\to 0$ limit the computed $n(r)$ is independent of system size, within statistical uncertainties. The greater the system size, the higher the temperature for which the computed $n(r)$ can be observed to plateau at large distances, i.e., for which an estimate of the condensate fraction $n_0(T)$ can be obtained.    }
}

@article{Boninsegni2023b,
   author = {Boninsegni, M.},
   journal = {Entropy},
   title = {{The Solid Phase of $^4$He: A Monte Carlo Simulation Study}},
   volume = {25},
   pages = {1114},
   number = {8},
   ISBN = {1099-4300},
   DOI = {10.3390/e25081114},
   year = {2023},
   type = {Electronic Article}
}

@article{Boninsegni2001,
  title = {Phase Separation in Mixtures of Hard Core Bosons},
  author = {Boninsegni, M.},
  journal = {Phys. Rev. Lett.},
  volume = {87},
  issue = {8},
  pages = {087201},
  numpages = {4},
  year = {2001},
  publisher = {American Physical Society},
  doi = {10.1103/PhysRevLett.87.087201},
  url = {https://link.aps.org/doi/10.1103/PhysRevLett.87.087201}
}

@article{Boninsegni2012b,
  title = {Population size bias in {Diffusion Monte Carlo}},
  author = {Boninsegni, M. and Moroni, S.},
  journal = {Phys. Rev. E},
  volume = {86},
  issue = {5},
  pages = {056712},
  numpages = {7},
  year = {2012},
  publisher = {American Physical Society},
  doi = {10.1103/PhysRevE.86.056712},
  url = {https://link.aps.org/doi/10.1103/PhysRevE.86.056712}
}

@article{Boninsegni2012d,
  title = {Role of Bose Statistics in Crystallization and Quantum Jamming},
  author = {Boninsegni, M. and Pollet, L. and Prokof'ev, N. and Svistunov, B.},
  journal = {Phys. Rev. Lett.},
  volume = {109},
  issue = {2},
  pages = {025302},
  numpages = {4},
  year = {2012},
  publisher = {American Physical Society},
  doi = {10.1103/PhysRevLett.109.025302},
  url = {https://link.aps.org/doi/10.1103/PhysRevLett.109.025302}
}

@article{Yu2024,
   author = {Yu, S. and Boninsegni, M.},
   title = {$^4${He} monolayer on graphene: a quantum {Monte Carlo} study},
   journal = {Commun. Theor. Phys.},
   volume = {76},
   number = {9},
   pages = {095701},
   DOI = {10.1088/1572-9494/ad5710},
   year = {2024}
}
%\begin{thebibliography}{99}
%\end{thebibliography}
\end{CJK*}
\end{document}